\documentclass[twocolumn,showpacs,preprintnumbers,amsmath,amssymb]{revtex4}
\usepackage{epsfig}
\usepackage{amsmath}
\usepackage{graphicx}
\usepackage{amsfonts}
\usepackage{amssymb}
\newcommand{\beq}{\begin{equation}}
\newcommand{\eeq}{\end{equation}}

\def\Tfrac#1#2{{#1}/{#2}}

\begin{document}

\title{Quasilinearization Method and  Summation of the WKB Series}

\author{R.~Krivec$^1$
       %\footnote{Electronic mail: rajmund.krivec@ijs.si}
       and
       V.~B.\ Mandelzweig$^2$}
       %\footnote{Electronic mail: victor@phys.huji.ac.il}}
\address{$^1$J. Stefan Institute, P.O.\ Box 3000, 1001 Ljubljana, Slovenia\\
         $^2$Racah Institute of Physics, Hebrew University, Jerusalem 91904, 
	 Israel}

\begin{abstract}
\smallskip 
Solutions obtained by the quasilinearization method (QLM) are compared with 
the WKB solutions. Expansion of the $p$-th QLM iterate in powers of $\hbar$ 
reproduces the structure of the WKB series generating an infinite number
 of the WKB terms with the first $2^p$ terms reproduced exactly. The QLM
 quantization condition leads to exact energies for the 
P\"{o}schl-Teller, Hulthen, Hylleraas, Morse, Eckart potentials etc.
For other, more complicated potentials the first QLM iterate, given by the closed
analytic expression, is  extremely accurate.
 The iterates converge very fast. The sixth iterate of the energy
for the anharmonic oscillator and for the two-body  Coulomb Dirac 
equation has an accuracy of 20 significant figures.
   
\end{abstract}
\maketitle

\section{Introduction} 
The quasilinearization method (QLM) and its iterations were constructed
\cite{K} 
as a generalization of the Newton-Raphson method \cite{CB} for
 the nonlinear differential equations to 
yield rapid quadratic and often monotonic convergence to the 
exact solution. It does not rely
on the existence of any kind of smallness parameter. The derivation 
of the WKB solution starts by casting the
Schr\"{o}dinger equation into nonlinear Riccati form and solving that
equation by expansion in powers of $\hbar$. It is interesting instead to
solve this nonlinear equation with the help of the quasilinearization
method (QLM) whose application to physical problems are discussed in 
works  \cite{VBM1,MT,KM1,RV} and compare with the WKB results.

The goal of this work is to point out that QLM iterates which are 
 expressible in a closed integral form provide better
approximation than the usual WKB. We show that  
the $p$-th QLM iterate  when expanded in powers of $\hbar$ reproduces
 the structure of the WKB series generating an infinite number
 of the WKB terms with the first $2^p$ terms reproduced exactly. A similar
 number of the next terms are reproduced with approximately correct
 coefficients. We prove also that the exact quantization condition 
in any QLM approximation, 
including the first, leads to exact energies for many known
  physical potentials used in molecular and nuclear physics such as 
the P\"{o}schl-Teller, Hulthen, Hylleraas, Morse, Eckart etc.
 In the general case of
arbitrary potentials that do not have a simple analytic structure, we 
illustrate that both the
wave functions and energies are very well reproduced by the first QLM
iterate and show significant improvement over those obtained by the usual
WKB approximation. If the initial QLM guess is
properly chosen, the wave function in the first QLM iteration, unlike the
WKB wave function, is free of unphysical turning point singularities. 
Since the first QLM iteration is given by an analytic expression
 \cite{VBM1,MT,KM1,RV}, it allows one to analytically
 estimate the role of
different parameters  and the influence of their variation on different
 characteristics of a quantum system. The next iterates display very
 fast quadratic 
convergence so that accuracy  of energies obtained after a few 
iterations is extremely high, reaching up to 20 significant figures 
for the sixth iterate 
as we show on the examples of the anharmonic oscillator and
the two-body Dirac equation with the Coulomb potential.

%RK%The usual WKB substitution $\chi(r)= C \exp\left(\lambda\int^r y(r') dr'\right)$ 
%RK%converts the Schr\"{o}dinger equation to nonlinear Riccati form 
The usual WKB substitution $\chi(r)=C\exp\left(\lambda\int^r y(r') dr'\right)$ 
converts the Schr\"{o}dinger equation to nonlinear Riccati form 
\beq 
\frac{dy(z)}{dz}+ \left(k^2(z)+y^2(z)\right)=0,
\label{eq:weq} 
\eeq 
%RK%Here $k^2(z)=E-V-\frac{l(l+1)}{z^2}$, $\lambda^2=\frac{2 m}{\hbar^2}$
where $k^2(z)=E-V-\Tfrac{l(l+1)}{z^2}$, $\lambda^2=\Tfrac{2 m}{\hbar^2}$,
 $z=\lambda r$.

The proper bound state boundary condition for potentials falling off at $z
\simeq z_0 \simeq \infty$ is $y(z)= \mathrm{const}$ at $z \geq z_0$.
 This means that $y'(z_0) = 0$, so that Eq.\ (\ref{eq:weq}) at 
$z \simeq z_0$ reduces to
$k(z_0)^2+y^2(z_0))=0$ or $y(z_0))= \pm i k(z_0)$. We choose here  
to define the boundary condition with the plus sign, so that 
$y(z_0)=  i k(z_0)$.

The
quasilinearization \cite{VBM1,MT,RV} of this equation gives a set of 
recurrence differential equations 
\beq
\frac{dy_{p}(z)}{dz}=y_{p-1}^2(z)-2 y_{p}(z)y_{p-1}(z)-k^2(z). 
\label{eq:qeq} 
\eeq
with the boundary condition  $y_{p}(z_0)= i k(z_0)$.
 
The analytic solution \cite{RV} of these equations expresses the $p$-th
iterate $ y_{p}(z)$ in terms of the previous iterate:
%RK%\begin{eqnarray}
%RK%y_{p}(z)& =& f_{p-1}(z)-\int_{z_0}^{z}ds \frac{d\,f_{p-1}(s)}{ds}\;
%RK%\exp[-2 \int_{s}^{z} y_{p-1}(t) dt],\nonumber \\
%RK% f_{p-1}(z)& =& 
%RK%\frac{y_{p-1}^2(z)-k^2(z)}{2 y_{p-1}(z)}.
%RK%\label{eq:ipqeq}
%RK%\end{eqnarray}
\begin{eqnarray}
y_{p}(z)& =& f_{p-1}(z)-\nonumber \\
&& \int_{z_0}^{z}ds \frac{df_{p-1}(s)}{ds}
\exp\left[-2\int_{s}^{z} y_{p-1}(t) dt\right],\nonumber \\
 f_{p-1}(z)& =& 
\frac{y_{p-1}^2(z)-k^2(z)}{2 y_{p-1}(z)}.
\label{eq:ipqeq}
\end{eqnarray}
Indeed, differentiation of both parts of Eq.\ (\ref{eq:ipqeq}) leads
 immediately to 
Eq.\ (\ref{eq:qeq}) which proves that  $y_{p}(z)$ is a solution of 
this equation. The boundary condition is obviously 
satisfied automatically.

The successive integrations by parts of Eq.\ (\ref{eq:ipqeq}) 
lead \cite{RV} to the series
\beq
y_{p}(z)= \sum_{n=0}^{\infty}\mathcal{L}_n^{(p)}(z)
\label{eq:piqeq}
\eeq
with $\mathcal{L}_n^{(p)}(z)$ given by recursion relation
%RK%$\mathcal{L}_n^{(p)}(z)=\frac{1}{2\,y_{p-1}(z)}\frac{d}{dz}
%RK%(-\mathcal{L}_{n-1}^{(p)}(z)),\,\, \mathcal{L}_n^{(0)}(z)=f_{p-1}(z)$.
$\mathcal{L}_n^{(p)}(z)=\Tfrac{1}{(2y_{p-1}(z))}\,\Tfrac{d%
(-\mathcal{L}_{n-1}^{(p)}(z))}{dz},\,\, \mathcal{L}_n^{(0)}(z)=f_{p-1}(z)$.
Since 
%RK%$\frac{d}{dz}=g\frac{d}{dr},\;\;g=\lambda^{-1}=\frac{\hbar}{\sqrt{2 m}}$, 
$\Tfrac{d}{dz}=g\,\Tfrac{d}{dr},\, g=\lambda^{-1}=\Tfrac{\hbar}{\sqrt{2m}}$, 
Eq.\ (\ref{eq:piqeq}) represents the expansion of the $p$-th QLM iterate in 
powers of $g$,  
that is in powers of $\hbar$, which one can compare with the WKB series.
For the zeroth iterate $y_{0}(z)$ it seems natural to choose the zeroth WKB
approximation, that is to set 
$y_{0}(z)=i k(z)$, 
which in addition automatically satisfies the boundary condition.
However, one has to be aware that this choice has unphysical turning point
singularities. According to the existence theorem for linear
differential equations \cite{In}, if $y_{p-1}(z)$ in Eq.\ (\ref{eq:qeq}) 
is a discontinuous function of $z$ in a certain interval, then 
$y_{p}(z)$ or its derivatives may also be discontinuous functions in
this interval, so consequently the turning point singularities 
of $y_{0}(z)$ may propagate to the next iterates.

Eq.\ (\ref{eq:ipqeq}) gives an especially simple expression \cite{RV}
 for the first QLM iterate
$y_{1}(z) = ik(z) - i\int_{z_0}^{z} ds\, 
k'(s) \exp\left[-2i\int_{s}^{z} k(t) dt\right]$
which thus is expressible in a 
closed integral form. This expression takes 
 into account, though approximately, an infinite number of the WKB
 terms corresponding to higher powers of $\hbar$, as it will be shown 
 in the next section. In view of this it is a better 
 approximation than the usual WKB. 

To obtain the WKB series one has to expand the solution $y$ of the Riccati
equation (\ref{eq:weq}) in powers of $\hbar$. This is easy to do 
by looking for $y$ in the form of a series and equating terms  of
identical powers of $g$,
\begin{eqnarray} 
y&=&\sum_{m=0}^{\infty} g^m Y_m,\; \nonumber \\
Y_{m}&=&-\frac{1}{2\,Y_0}\,\left(Y'_{m-1}+\sum_{k=1}^{m-1} 
Y_k\;Y_{m-k}\right).
\label{eq:sumwkb}
\end{eqnarray}
The derivatives in this and subsequent expressions are in the variable $r$.
The zeroth WKB approximation $Y_0$ is given by $Y_0=ik$. 
The comparison of expansion of the first QLM iterate 
in powers of $\hbar$ and the WKB series was originally 
performed in \cite{RV}. There it was shown that 
the expansion
reproduces exactly the first two terms 
and also gives correctly the structure of the WKB series up to 
the power $g^3$ considered in these works, generating  series with 
proper WKB terms, but with different coefficients.  
Comparison of Eqs.\ (\ref{eq:piqeq})
and (\ref{eq:sumwkb}) in the present work shows that this 
conclusion is true also if higher powers of $g$ are taken into account.

The computation of the expansion of the second QLM iterate $y_{2}$
 in powers of $\hbar$
is done by reexpanding the term $\Tfrac{1}{(2\,y_{1})}$ 
in $\mathcal{L}_n^{(p)}(z)$ in the series in powers of $g$ 
and keeping the powers up to $g^7$ inclusively in 
this expression as well as in the sum in Eq.\ (\ref{eq:piqeq}).
This procedure performed with the help of Mathematica \cite{Math} shows that
 expansion of $y_2$ reproduces exactly already the first four terms 
of the WKB series. It also gives  the true structure of 
the next terms of the WKB series, generating  series with 
proper WKB terms which have approximately correct coefficients. 
 The expansion of $y_3$ is obtained in the similar fashion.
 It reproduces exactly the first eight terms of the WKB series,
 that is all the terms up to the power $g^7$ inclusively. 
 
 Summing up, we have proved that the expansion of the 
 first, second and third QLM iterates reproduces exactly 
 two, four and eight WKB terms respectively. Since the 
 zeroth QLM iterate $y_0$ was chosen to be equal to the zeroth WKB
 approximation $ik$, one can state that the $p$-th QLM iterate
 contains $2^p$ exact terms. In addition the expansion of each QLM 
 iterate has the proper structure  whose 
terms are identical to those of the WKB series but have only 
approximately correct coefficients.

The  $2^p$ law is, of course, not accidental. The QLM iterates
are quadratically convergent \cite{K, VBM1}, that is the 
norm of the difference of the
exact solution and the $p$-th QLM iterate $\|y-y_p\|$ is proportional 
to the square
of the norm of the differences of the  exact solution and 
the $(p-1)$-th QLM iterate:
\beq
\|y-y_p\| \sim {\|y-y_{p-1}\|}^2.
\label{eq:norm}
\eeq
Here the norm $\|g\|$ of the function $g(x)$ 
is the maximum of the function $g(x)$ on the whole interval 
of values of $x$.
Since $y_0$ contains just one correct WKB term of power $g^0$
 and thus
$\|y-y_0\|$ is proportional to $g$, one has  to expect 
that $\|y-y_1\| \sim g^2$
and thus $y_1$ contains two correct WKB terms of powers 
$g^0$ and $g^1$. The difference $\|y-y_2\| \sim \|y-y_1\|\sim g^4$ 
so that $y_2$ contains four correct WKB terms of 
powers $g^0$, $g^1$, $g^2$  and $g^3$. Finally, the difference
$\|y-y_3\|$ should be proportional to $g^8$, and therefore
$y_3$ has to contain eight correct terms with powers 
between $g^0$ and $g^7$ inclusively. This explains the $2^p$ law.

The exact quantum mechanical quantization condition for 
the energy \cite{D,LP1} has the form: 

\begin{equation}
J = {\oint}_C y(z)dz =i\,2 \pi n.
\label{eq1}
\end{equation} 
 Here $y(z)$ is the logarithmic derivative of the wave function, given by  Eq.\ 
(\ref{eq:weq}), $z = gr$,  $n = 0,1,2,...$ is the
bound state number which counts the number of poles
 of $y(z)$; the integration is along a path C in the complex
plane encircling the segment of the $\Re z$ axis between 
the turning points.

 The $p$-th QLM iterate $y_p(z)$, as we have seen, contains, in addition
to $2^p$ exact WKB terms of powers $g^0$, $g^1$, ${\ldots}$, $g^{2^p-1}$,
 also an infinite number of structurally 
correct WKB terms of higher powers of $g$ with approximate coefficients.
 One can expect therefore that the quantization condition (\ref{eq1})
 with $y(z)$ approximated by any QLM iterate $y_p(z),\; p=1,2, {\ldots}$, 
 including the first,
gives more accurate energy than the usual WKB quantization condition
 which  is obtained by substituting into
 exact quantization condition (\ref{eq1})
the WKB expansion up to the first power
 of $g \sim \hbar$,
that is $y(z)= ik(z) - \Tfrac{(\Tfrac{d k(z)}{dz})}{(2 k(z))}$,
and neglecting all higher powers of $g$ in the expansion.
Indeed, we will prove now  that Eq.\ (\ref{eq1}) with $y(z)$ 
approximated by any QLM iterate leads to exact 
energies not only for the Coulomb and harmonic oscillator potentials
as it was shown earlier in Ref.\ \cite{RV}, but for many other well 
known  physical potentials used in molecular and nuclear physics such as 
the P\"{o}schl-Teller, Hulthen, Hylleraas, Morse, Eckart etc.  
The WKB quantization condition yields the exact energy
only for the first two potentials, but not for the rest of them.

Let us prove it on the example of the Hulthen potential
$ V(r) = -A\Tfrac{\exp(-\Tfrac{r}{a})}{(1-\exp{(-\Tfrac{r}{a})})}, 
A > 0,\, 0 < r < \infty$,
which plays an important role in molecular and nuclear physics.

To compute energy levels in the quasilinear approximation we have to use 
the  QLM equation (\ref{eq:qeq}) which after switching to the variable 
$t=\exp{(-\Tfrac{r}{a})}, 0<t<1$ has the form
\begin{eqnarray}
- \frac{t}{a} \frac{dy_p}{dt} = y^2_p -2 y_p\,y_{p-1}
 +\epsilon - A \frac{t}{1-t} 
\label{eq:ecenqlm}
\end{eqnarray}
Here $\epsilon=-E$ and $E$ is the energy. For convenience of 
further computations we set here $\Tfrac{2m}{\hbar^2}$ 
equal to unity. 
 The quantization condition (\ref{eq1}) in variable $t$ is given by
$J_p =a {\oint}_C\,(\Tfrac{y_p(t)}{t})\, dt =i2\pi n$, $p=1,2, {\ldots}$
 
 At the singular point $t \sim 0 $ of the integrand,
 Eq.\ (\ref{eq:ecenqlm}) reduces to
$-(\Tfrac{t}{a}) (\Tfrac{dy_p}{dt}) = y^2_p -2 y_p\,y_{p-1}+\epsilon$
whose solution is $ y_p = c_p$, where  $c_p$ is a constant satisfying
the algebraic equation $c^2_p -2 c_p\,c_{p-1}+\epsilon = 0$. Since at
large $p$  we expect $y_{p-1} \rightarrow y_p \rightarrow y$, where 
$y$ is an exact solution at $t=0$, it 
means, in view of $c_p$ being a constant, that we should have
$c_{p} = c_{p-1}=c$, that is, we are looking for a fixed point 
solution of this algebraic equation, which is $c_p=\sqrt{\epsilon}$.
The positive sign before the root is chosen since  
the first term in expansion of $y_p(t)$ in the WKB terms is $ik(t)$. 
Thus  $y_p(t) \simeq i k(t)= i \sqrt{-\epsilon+ 
 \,A\Tfrac{t}{(1-t)}}$, that is  $y_p(0) \simeq +\sqrt{\epsilon} $. 

In the same way one finds that the residues at other singular points $t=1$
and  $t=\infty$ equal $\Tfrac{1}{a}$ and $\sqrt{\epsilon +A}$, respectively.  
After taking into account that poles $t=0$ and $t=1$ are encircled in the 
opposite direction compared with the pole at infinity (which is
encircled counterclockwise) and the reinstatement of the factor 
$\Tfrac{2m}{\hbar^2}\,$  the exact quantization condition 
gives $\sqrt{\Tfrac{2m a^2}{\hbar^2}} (-\sqrt{\epsilon}+
\sqrt{\epsilon+A})= n+1,\,\, n=0,1,2, {\ldots}$. 
This expression coincides with the exact one for the Hulthen
potential calculated in \cite{F,G} and is different from 
the WKB quantization condition $\sqrt{\Tfrac{2m a^2}{\hbar^2}} 
(-\sqrt{\epsilon}+\sqrt{\epsilon+A}) = n+\frac {1}{2}$. 

Similar computations show that the quantization  
condition (\ref{eq1}) in any QLM approximation including the first
 leads to exact 
energies for all the potentials mentioned above and for 
other potentials with a simple singular structure.

For more complicated potentials numerical calculation is necessary. 
 However, as we will see now, already the first QLM iterate, given 
 by the closed analytic expression, is  extremely accurate. 
 For the zeroth iterate $y_{0}(z)$ one can choose the usual WKB
approximation. However, this choice has unphysical turning point
singularities. Consequently, if $y_{p}(z)$ in Eq.\ (\ref{eq:qeq}) is a
discontinuous function of $z$ in a certain interval, then \cite{In}
$y_{p+1}(z)$ or its derivatives could also be discontinuous functions in
this interval, so the turning point singularities of $y_{0}(z)$ will
unfortunately propagate to the next iterate.
 To avoid this we choose the
Langer WKB wave function \cite{Lan} as the zeroth iteration. This function
near the turning points $a$ and $b$ is given by the simple analytic
expression 
\begin{eqnarray}
\chi_i(r)&=&c_i \sqrt{\frac{S_i^{\frac{1}{3}}(r)}{\left|k(r)\right|}}
\mathrm{Ai} \left[d\; S_i^{\frac{1}{3}}(r)\right], \nonumber\\ 
S_i(r)&=&\frac{3}{2} \lambda
\left| \int_{i}^{r} \left| k(s)\right| ds \right|.
\label{eq:lan}
\end{eqnarray}
Here Ai denotes the Airy function, $i=a,b$,  
$k(r)=(\Tfrac{2 m}{\hbar^2}) 
\left[(E-V(r)) - \Tfrac{(l+\Tfrac{1}{2})^2}{(2mr^2)}\right]$,
  $d$ is $-1$ for $a<r<b$
and $1$ for $r \leq a, r \geq b$, and $c_a=1$, $c_b=(-1)^n$, where
$n=0,1,2, {\ldots}$ is the number of the bound state. It is easy to check 
that $\chi_a(r)$ and $\chi_b(r)$ coincide at some point at the interval
$(a,b)$ between the turning points, are
continuous across them and coincide with 
the usual WKB solutions far from them.
 
Let us consider a couple of examples. The exact energy of
 the ground state of the anharmonic oscillator $V(r) = r^4$
  is 2.393\;644\;016\;482\;303\;115\;6  in atomic units with mass set to unity, $m =
1$. This result is obtained by us using 
the Runge-Kutta method and quadruple precision arithmetic.
The WKB energy is different by 2.8\% and equals 2.32662, while 
the first-iteration QLM energy equals 2.39475
 and differs from the exact energy only by 0.046\%. The 
QLM energy coincides with the exact energy in all twenty digits 
after the sixth iteration.

\begin{figure}
\begin{center}
\epsfig{file=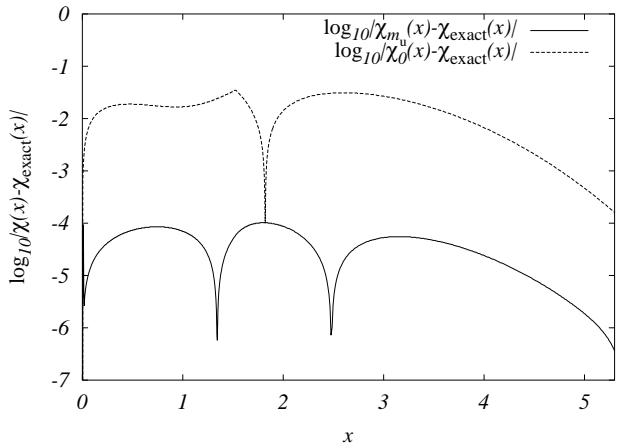,width=87mm}
\end{center}
\caption{Logarithm of the difference between the exact solution
$\chi_{\rm exact}$ and the WKB solution $\chi_{0}$
(dashed curve) and between $\chi_{\rm exact}$ and the first QLM
 iterate $\chi_{m_u}$ (solid curve) for the ground state of the anharmonic
 oscillator \cite{RF}.}
\label{fig2}
\end{figure}

The graph in Fig.\ \ref{fig2} displays 
the logarithm of the difference between the exact and WKB solutions and
between the exact solution and the first QLM iteration. One can
see that the difference between the exact solution and the first QLM
iteration is two orders of magnitude smaller than the difference between the
exact and the WKB solutions, that is, just one QLM iteration increases the
accuracy of the wave function by two orders of magnitude \cite{RF}.

The second example is the modified Coulomb potential 
\beq
V(r)= - \frac{1}{2 \rho} + \frac{l (l+1) - \frac{1}{4} \alpha^2}{\rho^2} + 
\frac{\frac{3}{4} \alpha^2}{\rho^2 (\rho +\alpha^2)^2},\; \rho = \alpha E r
\nonumber
\eeq
obtained when the equal-mass two-body Dirac equation
 with the static  Coulomb 
interaction is reduced to the Schr\"{o}dinger equation \cite{M}.
The exact energy of the $ ^1S_0$ ground state is 
0.999\;993\;340\;148\;538\;880\;12 in atomic units with double mass set to unity,
 $2M = 1$. This result was obtained in the work  \cite{S} by an elaborate 
computation using the finite element method and verified by ourselves
using the Runge-Kutta method in quadruple precision. 
 WKB in this case  predicts 
the energy very accurately since the potential is very close to the
Coulomb potential, for which WKB energy is precise. It equals 
0.999\;986\;680\;0 and differs
 from the exact one only by $6.66 \times 10^{-6}$. The
first-iteration QLM energy equals 0.999\;993\;335\;4
 and differs from  the exact energy by $5 \times 10^{-9}$,
that is, it is more accurate than the WKB energy by three orders of magnitude.
The QLM energy coincides with 
the exact one up to the twentieth digit after the sixth iteration. 

The calculation shows that the difference between the exact 
wave function and the first QLM
iteration is, as in previous example, by two orders of 
magnitude smaller
 than the difference between the exact and the WKB solutions. 
 Thus also in this case one QLM iteration
 increases the accuracy of the wave function by a remarkable 
two orders of magnitude.

In conclusion, we have shown that the quasilinearization method 
(QLM) which 
approaches solution of the Riccati-Schr\"{o}dinger equation 
 by approximating the nonlinear terms 
by a sequence of the linear ones, and is not based
on the existence of a small parameter, sums the WKB 
series.  The expansion  of the $p$-th QLM iteration
in powers of $\hbar$ reproduces the structure of the WKB 
series generating an infinite number of the WKB term 
with $2^p$ terms of the expansion reproduced 
exactly and a similar number approximately. As a result one expects
 that the exact quantization condition with integrand
replaced by any QLM iterate including the first
gives more accurate energy than the WKB quantization condition
 which  is obtained by substituting into Eq.\ (\ref{eq1}) of 
the WKB expansion up to the first power of $\hbar$
and neglecting all higher powers of $\hbar$. 
We show on examples of the Hulthen potential
that it in fact, the QLM energy is exact already 
in the first iteration. Similarly, one can show that the approximation
 by the first QLM iterate in Eq.\ (\ref{eq1}) leads to exact energies
 for many well 
known  physical potentials such as the Coulomb, harmonic 
oscillator, P\"{o}schl-Teller, Hulthen, Hylleraas, Morse, Eckart etc. 
For other potentials which have more complicated analytical structure
we show on examples of the anharmonic oscillator and modified Coulomb 
potentials that the use of the Langer WKB solution as an
initial guess already in the first QLM approximation gives energies and wave
functions at least two orders of magnitude more accurate than the WKB results.
Such a QLM solution, unlike the usual WKB solution, displays no unphysical
turning point singularities. Since the first QLM iterate is given by
 a close analytic
expression it allows one to estimate analytically the role of
different parameters and their influence on properties of a quantum system
 with much higher precision than provided by the WKB approximation. 
In addition, it was shown that six QLM iterations are
usually enough to obtain both the wave function and the energy with extreme
accuracy of twenty significant figures.

The research was supported by the Bilateral Cooperation Program at
the Ministry of Education, Science and Sport of Slovenia (RK) and by
the Israeli Science Foundation grant 131/00 (VBM).


\begin{references}
\bibitem{RF}{The dips are artifacts of the logarithmic
scale, as the logarithm of the absolute value of the difference of two 
solutions goes to $- \infty$ at points where the difference changes
sign. The overall accuracy of the solution can therefore be inferred only
at $x$ values not too close to the dips.}
\end{references}
\end{document}